# Metastable magnetic domain wall dynamics

**Mahdi Jamali[1], Kyung-Jin Lee[2] and Hyunsoo Yang[1,3]**

[1] Department of Electrical and Computer Engineering, National University of Singapore, 4 Engineering Drive 3, Singapore 117576, Singapore
[2] Department of Materials Science and Engineering, Korea University, Seoul 136-713, Republic of Korea
[*] E-mail: eleyang@nus.edu.sg

**Abstract.** The dynamics of metastable magnetic domain walls in straight ferromagnetic nanowires under spin waves, external magnetic fields, and current induced spin transfer torque are studied by micromagnetic simulations. It is found that in contrast to a stable wall, it is possible to displace a metastable domain wall in the absence of any external excitation. In addition, independent of the domain wall excitation method, the velocity of a metastable wall is much smaller than a stable wall and their displacement direction could be different from the stable wall depending on the structure of metastable walls. Under the current induced spin transfer torque excitation, the direction of domain wall displacement is directly related to the intensity of nonadiabatic spin transfer torque. In a rough nanowire, it is found that the displacement of a metastable wall could happen much below the critical excitation of a stable wall. Furthermore, we show that it is possible to have either a forward or backward displacement of a metastable domain wall by changing the pulse width of the excitation.

[3] Author to whom any correspondence should be addressed.

## 1. Introduction

The controlled displacement of magnetic domain walls by current induced spin transfer torque in a ferromagnetic nanowire has been intensively studied recently due to its potential applications for next generation solid state memories[1-3] and logic devices[4, 5]. In a ferromagnetic nanowire various types of domain walls with different energy states can exist. Typically only one structure is in the global energy minimum state and forms a stable domain wall, while the other structures form metastable walls depending on the width and thickness of a ferromagnetic nanowire [6, 7]. However, it has been shown experimentally and by micromagnetic simulations that the phase diagram of domain walls could be changed, when the domain wall is nucleated by application of a transverse field [8, 9]. In addition, at an elevated temperature, a domain wall structure can transform to other structures [10-12]. The thermal effect due to a high current density required for the domain wall displacement can also change the domain wall structure and nucleate a metastable domain wall [13-18].

The displacement of a magnetic domain wall under a magnetic field is expected to only depend on the direction of the magnetic field and whether the domain wall is head to head or tail or tail, and be independent of domain wall structures. In addition, the domain wall motion induced by spin trasnfer torque is anticipated to be related with the direction of conduction electron flow. Although most of experimental results have shown the magnetic domain wall propagates in the direction of electron flow [14, 19-26], there have been some reports that have observed bi-directional domain wall motion in the presence of unipolar current pulses [27, 28].

In this report, we have correlated the bi-directional motion to the dynamics of metastable domain walls. We have studied the dynamics of metastable domain walls under different types of excitations and compared with that of a stable domain wall. It is found that under the spin wave excitation, a stable wall always moves in the propagation direction of spin waves whereas a metastable wall could displace in the reverse direction until it transforms to a stable wall. In addition, the velocity of a metastalbe wall is much lower than that of a stable wall. In the case of current induced domain wall motion, the direction of the metastable wall displacement is strongly related to the nonadiabaticity of spin transfer torque. In a rough nanowire, it is found that the metastable wall could have a finite displacement in a magnetic field or current much below the critical field or current density required to displace a stable wall. In addition, depending on the structure of the metastable wall and the excitation pulse width, the metastable wall could have a birectional displacement under a unipolar excitation.

The structure that we have used in the simulations is shown in figure 1(a). The wire has a width of 100 nm, thickness of 10 nm, and length of 4 µm. The simulation cell size is 4×4×10 nm$^3$ and the nanowire is made of Permalloy (Py) with the saturation magnetization ($M_S$) of 860×10$^3$ A/m, the exchange stiffness ($A_{ex}$) of 1.3×10$^{-11}$ J/m, and the Gilbert damping constant ($\alpha$) of 0.01. We have located different kinds of domain walls in the nanowire and measured the total energy of the system. It is found that a transverse wall has the minimum energy of 1.268×10$^{-17}$ J while the total energy of the nanowire in the presence of a vortex and an antivortex wall is 1.374×10$^{-17}$ J and 1.474×10$^{-17}$ J, respectively, as shown in figure 1(b). Although the total energy of the nanowire for the vortex and the antivortex are only 8.4% and 16.2%, respectively, larger than that of the transverse wall, we observe drastic changes in the domain wall dynamics in the presence of external excitations such as spin waves, external magnetic fields, and current induced spin transfer torque. A higher energy state of the nanowire in the presence of the metastable wall is associated to the energy that has stored inside the domain wall as anisotropy energy [29]. It is possible to release this stored anisotropy energy and transform a metastable wall to a stable wall by overcoming the energy barrier which is ΔE$_1$ between the antivortex and the transverse wall and ΔE$_2$ between the vortex and the transverse wall in figure 1(b). If the external excitation is strong enough to overcome this energy barrier, the stored energy in a domain wall can displace the domain wall in the either forward or backward direction even after removing the external excitation. It should be pointed out that reverse of this process happens for the excitation above the Walker breakdown that could transform a stable wall to a metastable wall [14, 30, 31]. We first present domain wall dynamics in a perfect nanowire and compare the results with the domain wall automotion equation. Then domain wall dynamics in a periodic and random rough nanowire are shown and compared with the case of a perfect nanowire.

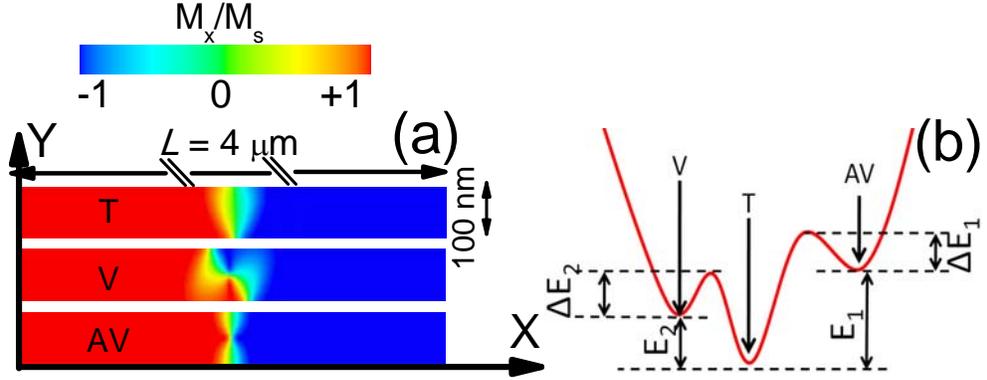

**Figure 1.** (a) Different types of domain walls [transverse (T), vortex (V), and antivortex (AV)] located at the center of the nanowire. (b) The energy landscape of the nanowire in the presence of different types of domain walls.

## 2. Metastable wall dynamics in perfect nanowires

*2.1. Spin wave excitation*

In order to generate spin waves, we applied an external magnetic field to the first $4\times100\times10$ nm$^3$ on the left side of the nanowire. The applied field varied sinusoidally in time with frequency ($f$) and amplitude $H_0$ as $H = H_0 \sin(2\pi t f)\hat{y}$ in the $y$-direction. We have used an absorbing boundary condition to prevent spin wave reflection from the edge of the nanowire [32-34]. Figure 2(a) shows the motion of an antivortex (metastable) wall under spin waves with $f = 16$ GHz and $H_0 = 3$ kOe. In contrast to the previous reports [33, 35, 36], the domain wall moves in the reverse direction (toward spin wave source) for 1.7 μm until its core transits the width of nanowire at 34 ns. The antivortex wall is then converted to a transverse wall which is the minimum energy state and moves in the forward direction (+x-direction). The domain wall displacement profile is shown for a transverse, a vortex, and an antivortex wall in figure 2(b). The velocity of a vortex wall is much smaller than that of a transverse wall, although both walls move in the same direction.

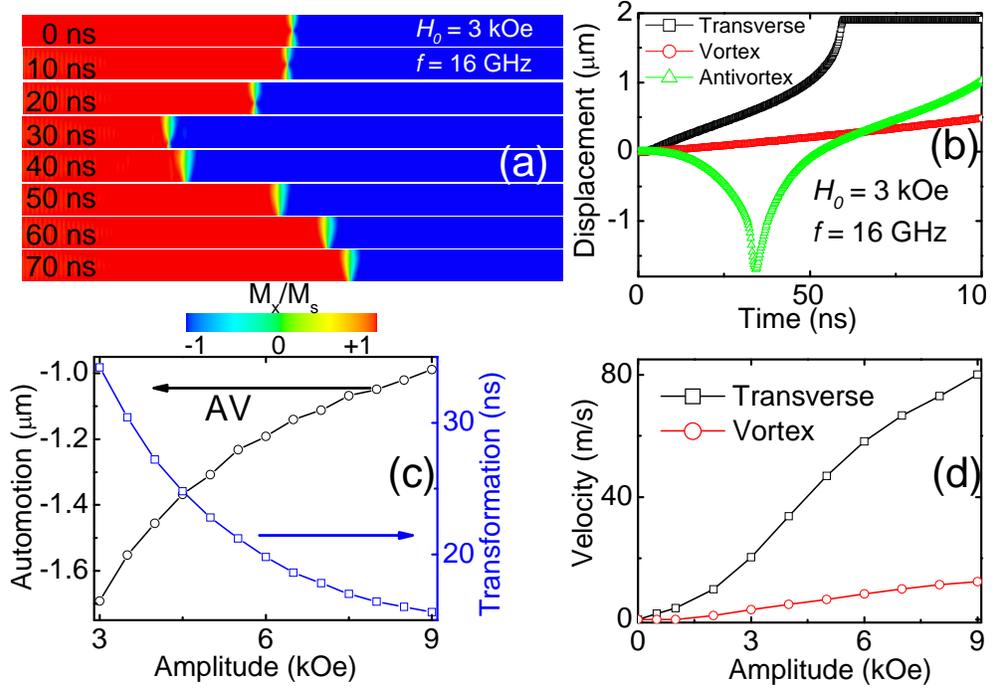

**Figure 2.** (a) The real time position of an antivortex wall for spin waves of $f = 16$ GHz and $H_0 = 3$ kOe. (b) Displacement profile of a transverse, vortex, and antivortex in the presence of spin waves with $f = 16$ GHz and $H_0 = 3$ kOe. (c) Antivortex automotion displacement and transformation time for spin waves of $f = 16$ GHz and different excitation amplitudes. (d) Linear velocity of a transverse and vortex wall for spin waves of $f = 16$ GHz and different excitation amplitudes.

We have used different excitation amplitudes and calculated the amount of reverse displacement and the required time for the transformation of an antivortex to a transverse wall in figure 2(c). By increasing the excitation amplitude from 3 to 9 kOe at a constant frequency of 16 GHz, the reverse displacement decreases from 1.7 to 1 μm and the transformation time decreases from 34 to 15.6 ns. The only mechanism for the nucleation of a stable wall in the perfect nanowire is the external excitation, therefore, by increasing the excitation amplitude of spin waves, the antivortex could transform to a stable wall in less time. We also have simulated a transverse and a vortex wall under spin waves with a frequency of 16 GHz and different excitation amplitudes, and measured the domain wall linear velocity as shown in figure 2(d). The domain wall linear velocity has been determined based on the time required for the displacement of domain wall between 5 to 200 nm. It is found that they always move in the +x-direction in contrast to the case of an antivortex and the velocity of the vortex wall is about 7 times smaller than that of the transverse wall in all the excitation amplitudes.

Most of the magnetic domain wall properties can be explained based on the hamiltonian conjugate variables; the domain wall center position ($q$) and the domain wall tilting angle towards out of the plane of the center of the domain wall magnetization ($\phi$) [37-39]. The dynamics of metastable domain wall in a ferromagnetic nanowire could be understood from the domian wall automotion or domain wall streaming equation. For any types of the domain wall in a perfect nanowire the following equation holds [28, 37]:

$$\frac{d\phi}{dt} + \frac{\alpha}{\Delta_t}\frac{dq}{dt} = F_{ext} \tag{1}$$

In which $\alpha$ is the Gilbert damping constant, $\Delta_t$ is the Thiele domain wall width, and $F_{ext}$ represents any external excitation. As seen in Eq. (1), in the absence of the external excitation, any change in the domain wall tilting angle would result in a change in the domain wall position. For a vortex or an antivortex wall, the tilting angle is essentially determined by the domain wall core and by time integration, one can obtain:

$$[q] = \pm\frac{\Delta_t}{\alpha}\frac{\pi}{w}p[y_c] \tag{2}$$

Here $p$ is the vortex core polarity, $w$ is the width of the nanostrip, and $y_c$ is the transverse displacement of a vortex or an antivortex core in the y-direction. The direction of the displacement for the vortex and the antivortex are opposite [28, 40]. This is consistent with our findings for the spin wave excitation in a vortex and an antivortex wall. The maximum amount of the displacement in the nanowire due to the domain wall transformation based on equation (2) is expected to be around 1.9 μm for $\Delta_t$ = 12 nm which is very close to our results shown in figure 2(c) for the antivortex case driven by spin waves. A minimum external excitation is required to overcome the energy barrier $\Delta E_1$ or $\Delta E_2$ in figure 1(b) in order to observe the transformation of a vortex or an antivortex to a transverse wall. The value of $\Delta E_1$ or $\Delta E_2$ could be determined from the results of the magnetic field excitation.

*2.2. Magnetic field excitation*

Figure 3(a) shows the domain wall displacement profile for a transverse, a vortex, and an antivortex under a 10 Oe magnetic field excitation. Similar to the spin wave excitation, the antivortex starts to move toward the –x-direction, while the transverse and vortex walls move in the +x-direction. From equation (2), the displacement direction of domain wall automation is independent of the external excitation which is in line with our results. For the vortex wall, it moves 320 nm to the +x-direction for first 8.4 ns and then it moves back to 76 nm up to 9.5 ns till the vortex core transits the nanowire width and transforms to a transverse wall. For the antivortex case, it moves 670 nm in –x-direction for the first 6.6 ns, and it moves in +x-direction after the transformation to a transverse wall. The average domain wall velocity for a transverse wall at different values of the magnetic field is shown in figure 3(b). The Walker field is around 17 Oe for the transverse wall above which the velocity of domain wall drops from 402 m/s to less than 183 m/s. From the field excitation, the energy barrier between the antivortex and the transverse wall ($\Delta E_1$) is determined to be around 3 Oe and the energy barrier between the vortex and the transverse wall ($\Delta E_2$) is determined to be around 1.5 Oe. By increasing the excitation field from 5 to 20 Oe, we determine the automotion displacement and the transformation time from a vortex or an antivortex to a transverse wall. In the case of a vortex wall, it is found that the amount of net reverse displacement decreases from 440 to 16 nm and the transformation time also declines from 22.6 to 4.5 ns, when the field changes from 5 to 20 Oe as shown in figure 3(c). For an antivortex, the reverse displacement decreases from 1000 to 416 nm and the transformation time also reduces from 11.2 to 3.7 ns, when the field increases from 5 to 20 Oe in figure 3(d).

We have also simulated the response of a vortex and an antivortex wall for the pulse magnetic field of 10 Oe with the different pulse widths in figure 3(e) and 3(f), respectively. It is found that when the pulse width is less than the domain wall transformation time, the vortex or antivortex continues their displacement without any transformation until it reaches the end of the nanowire due to the finite moment of a domain wall. When the pulse width is longer than the domain wall transformation time which is 9.5 and 6.5 ns for the vortex and antivortex respectively, the vortex or antivortex does the transformation and then the transverse wall moves toward +x- direction.

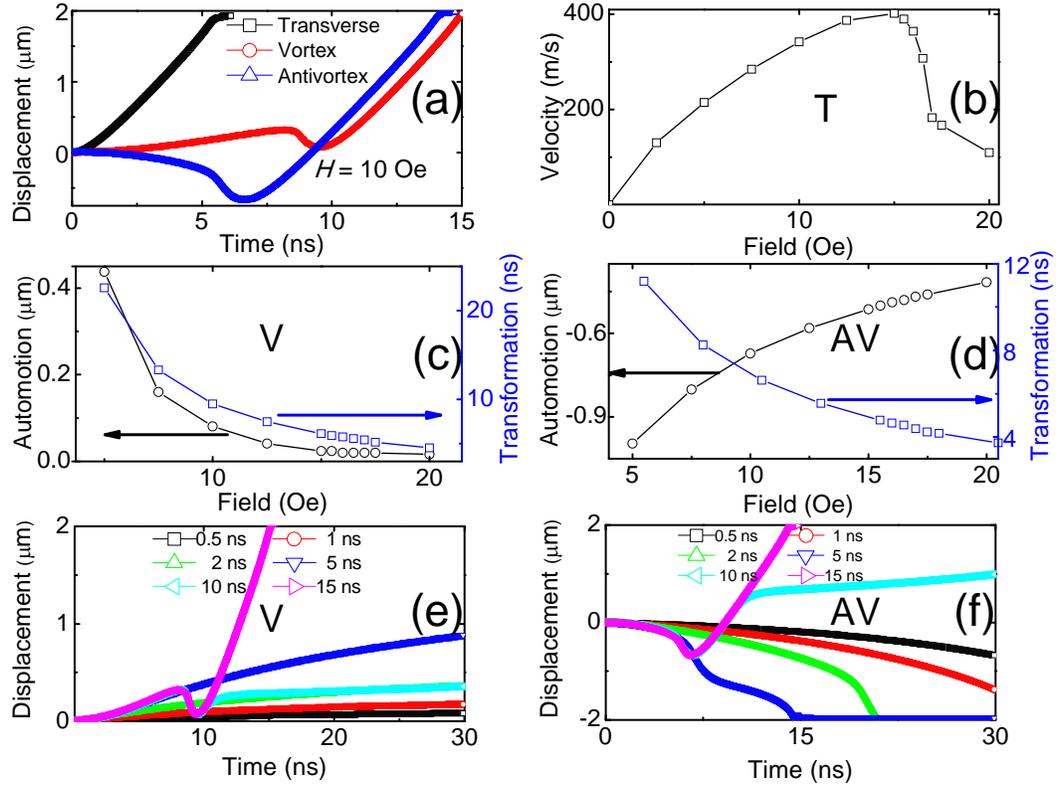

**Figure 3.** (a) Displacement profile of a transverse, vortex, and antivortex wall for a magnetic field of 10 Oe. (b) Average velocity of a transverse wall at different magnetic fields. Automotion displacement and transformation time of a vortex wall (c) and an antivortex wall (d) at different magnetic fields. A vortex (e) and an antivortex (f) wall displacement profile under a pulse magnetic field amplitude of 10 Oe and different pulse widths.

*2.3. Electric current excitation*

For the better understanding of the behavior of the metastable domain wall, we have used current driven spin transfer torque and simulated the response of different domain walls to spin transfer torque. Based on the domain wall one dimensional model for a transverse wall, the domain wall dynamics can be described as:

$$\dot{\phi} + \alpha \frac{\dot{q}}{\Delta} = \gamma_0 H_a + \frac{\beta u}{\Delta}, \quad \frac{\dot{q}}{\Delta} - \alpha \dot{\phi} = \gamma_0 H_k \sin(\phi)\cos(\phi) + \frac{u}{\Delta} \quad (3)$$

where $\gamma_0$ is the gyromagnetic ratio, $H_a$ is the applied magnetic field, $H_k$ is the transverse anisotropy, $\Delta$ is the domain wall width, $\beta$ characterizes the nonadiabatic contribution, and the $u$ parameter is the effective drift velocity of the conduction electron spins defined by $u = JPg\mu_B/(2eM_S)$, where $J$ is the current density, $P$ is the spin polarization, $\mu_B$ is the Bohr magnetron, and $e$ is the electron charge.

Figure 4(a) displays the response of a vortex wall to an electric current with $u = 50$ m/s ($\sim J = 1\times10^8$ A/cm$^2$) and different nonadiabatic coefficients. For small nonadiabatic coefficients ($\beta/\alpha < 2$), a vortex wall cannot overcome the energy barrier of $\Delta E_2$ and no transformation from a vortex to a transverse wall has been observed, whereas for larger value of $\beta$ ($\beta/\alpha \geq 2$) the transformation of a vortex to a transverse wall could happen and the transformation time decreases as the nonadiabatic spin transfer torque increases for the simulated time range. The behavior of an antivortex wall to the spin transfer torque excitation is also simulated in figure 4(b). In contrast to the magnetic field excitation, it is found that an antivortex wall first moves to the +x-direction similar to a vortex wall, however it moves to the –x-direction during its transformation. In addition, the transformation happens earlier for a larger value of nonadiabatic coefficient, which is similar to a vortex wall.

In order to understand this behavior, we should notice the difference between the spin transfer torque excitation and the magnetic field excitation. In principle, the magnetic field tilts the energy profile in figure 1(b) in such a way that a metastable domain wall could easily overcome the energy barrier and transform to a stable wall. The nonadiabatic spin transfer torque is similar to an effective magnetic field of $H_{eff} = \frac{\beta u}{\Delta \gamma_0}$ as can be seen from equation (3), and an enough value of the nonadiabatic torque is required to overcome the energy barriers and a larger nonadiabatic value decreases the transformation time of a metastable wall. For example, the value of an effective field originated from nonadiabatic spin transfer torque is around 11.8 Oe for $\beta$ = 0.05, $u$ = 50 m/s, and a domain wall width of $\Delta$ = 12 nm.

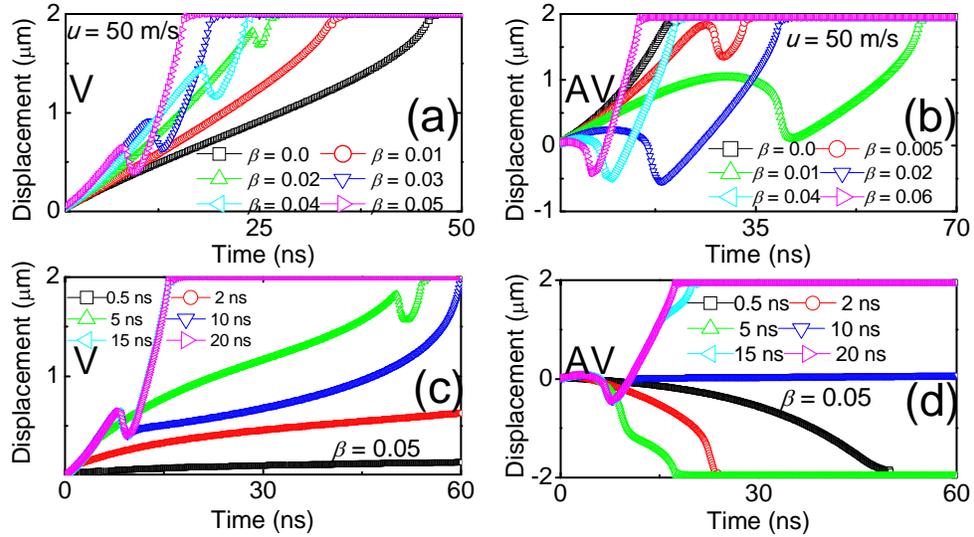

**Figure 4.** Displacement profile of a vortex (a) and an antivortex (b) wall under currents of $u$ = 50 m/s and different nonadiabatic coefficients. Displacement profile of a vortex (c) and an antivortex (d) under currents of $u$ = 50 m/s and $\beta$ = 0.05 for different current pulse widths.

We have also modeled the current pulse response of a vortex wall for the different pulse widths of current excitations and the nonadiabatic coefficient of $\beta$ = 0.05 in figure 4(c). Similar to the magnetic pulse response, if the excitation pulse is less than the required time for the vortex transformation, the vortex wall preserves its structures and moves up to the end of the nanowire, whereas for the longer current pulse, the vortex wall transforms to a transverse wall. The same behavior is observed for the antivortex response to a current pulse in figure 4(d). For a short pulse (< 5 ns), the antivortex wall displaces 90 nm toward the +x-direction due to adiabatic spin transfer torque, however, after removing the current pulse it moves to the –x-direction due to domain wall automation according to equation (2). If the excitation pulse is long enough (≥ 15 ns) so that the antivortex wall could transform to a transverse wall, the domain wall eventually moves toward +x- direction due to finite moment of transverse wall.

## 3. Metastable wall dynamics in rough nanowire

*3.1. Periodic roughness*

It is well known that the presence of roughness in a ferromagnetic nanowire could help a nucleation of domain wall [31, 41, 42]. In addition, a fabricated ferromagnetic nanowire always has a finite edge roughness and it would be interesting to see the roughness effect on the metastable domain wall dynamics. We have introduced a periodic roughness with a periodicity of $T = 25$ nm and depth of $D = 4$ nm into the nanowire as shown in the inset of figure 5(b). The periodicity of roughness is selected to be comparable to the width of domain wall and the chosen depth of roughness gives a depinning field of ~20 Oe for a transverse domain wall, which is close to the experimental reports [43, 44].

*3.1.1. Magnetic field excitation.* Figure 5(a) shows the displacement of a stable transverse wall for the different external magnetic fields. As can be seen, the transverse wall displacement is almost zero below 22 Oe. We have also calculated the average velocity of a transverse wall at the different excitation fields in figure 5(b). The average velocity of the transverse wall is very small (1.5 m/s) upto 20 Oe and the velocity suddenly jumps to 360 m/s at 22 Oe. In addition, the Walker field has been shifted from ~16 to 36 Oe as expected [31].

The response of a vortex wall to the different values of magnetic fields in a rough nanowire is shown in figure 5(c). In contrast to the transverse wall, we observe a finite displacement of the vortex wall below the depinning field of a stable (transverse) wall. Due to the domain wall automation, the stored energy inside the domain wall could assist the displacement of the domain wall, however, after the vortex wall transforms to a stable wall (a transverse wall), it stops due to pinning of roughness, unlike to the case of a perfect nanowire in which it reaches to the end of the nanowire. By increasing the magnetic field, the forward displacement of the vortex wall increases but it is not monotonic. This behavior could be explained based on the fact that in a rough nanowire there is a trade-off between the metastable automation and depinning of domain wall. By increasing the field, the depinning process of metastable wall would be assisted, while the automotion of the metastable wall would be limited, therefore, in a certain value of magnetic field, one can expect to observe the maximum displacement of the domain wall, which is 10 Oe in our nanowire.

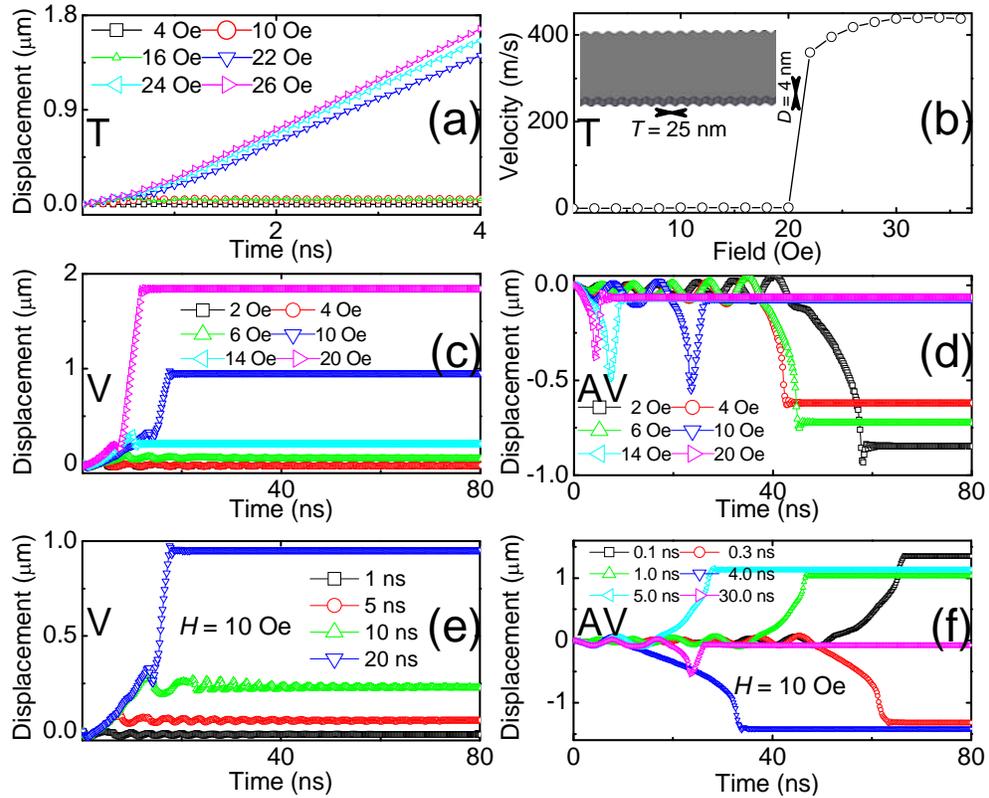

**Figure 5.** Displacement profile (a) and average velocity (b) of a transverse wall in a rough nanowire at different magnetic fields. Displacement profile of a vortex (c) and an antivortex (d) wall in a rough nanowire at different excitation fields. Displacement profile of a vortex (e) and an antivortex (f) wall in a rough nanowire for a pulse field of 10 Oe and different pulse widths.

We have also studied the displacement profile of an antivortex wall for the different values of magnetic fields in figure 5(d). By increasing the magnetic field, the domain wall transient time decreases as the field assists the domain wall depinning process. Furthermore, for small values of the field the mechanism of depinning is accompanied by resonant depinning of the antivortex [45, 46]. The roughness can create a local harmonic potential profile for an antivortex wall and it can help a conversion between the kinetic and potential energy of an antivortex wall. The antivortex wall oscillates between the pinning cites till its energy is large enough to overcome the energy barrier.

In figure 5(e), the response of a vortex wall to a pulse magnetic field is demonstrated for the different pulse widths and a field amplitude of 10 Oe. As can be seen, by increasing the pulse width, the vortex wall displacement increases as the external field helps the vortex wall to overcome the pinning sites. The pulse field response of an antivortex wall at a field amplitude of 10 Oe is shown in figure 5(f). Due to resonance depinning of the antivortex wall, it is possible to have a displacement in either forward or backward direction depending on the phase difference between the pulse magnetic field and resultant transient response of a domain wall (below 40 ns with 2 Oe). The displacement is almost independent of the pulse width, which is about ± 1.2 – 1.5 μm slightly smaller than the automotion displacement (1.9 μm) in a perfect nanowire. It is interesting to compare the dc field excitation of an antivortex wall at 10 Oe in figure 5(d) and its short pulse excitation in figure 5(f). In the dc field excitation, the antivortex has a finite displacement of ~ 70 nm in the –x-direction, while it can have a much longer displacement in the both +x and –x-directions with a short pulse of magnetic field. If the pulse width of field is longer than ~ 25 ns (the transformation time of the antivortex with 10 Oe), the bi-directional displacement would be suppressed.

*3.1.2. Electric current excitation.* We have also simulated the current effect on the wall dynamics in a rough nanowire. In figure 6(a), the displacement of a transverse wall at the different values of $u$ is shown for a nonadiabatic coefficient of $\beta$ = 0.05. For the $u$ values below 150 m/s, the transverse wall displacement is negligible, whereas for values of $u$ above 160 m/s domain wall transits to the end of the nanowire. In figure 6(b), we have measured the critical current density for depinning of the transverse wall at the different value of nonadiabatic coefficients. By decreasing the nonadiabaticity of the current, the critical current density increases, which is in line with a previous study [9].

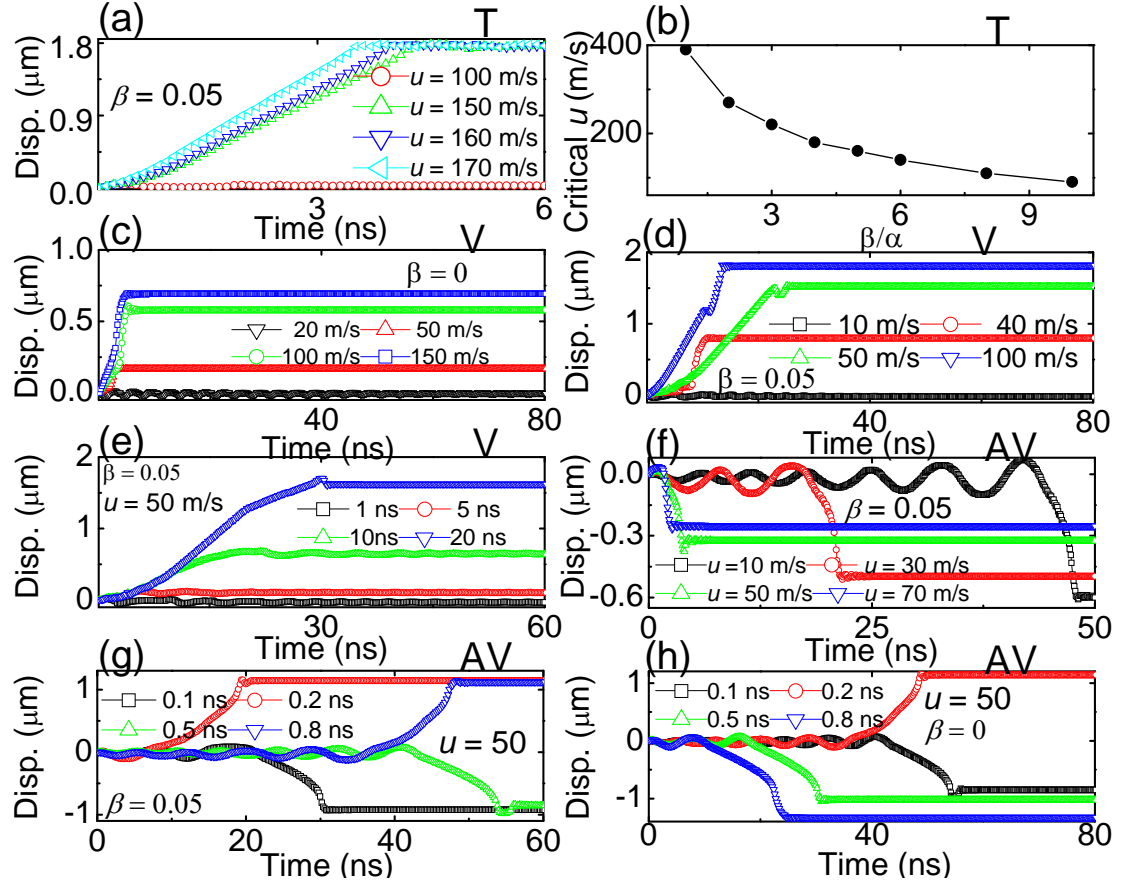

**Figure 6.** (a) Displacement profile of a transverse wall at different current densities for $\beta$ = 0.05. (b) Critical current density required for depinning of a transverse wall at different nonadiabatic coefficients. Displacement profile of a vortex wall at different current densities for $\beta$ = 0 (c) and $\beta$ = 0.05 (d). (e) Displacement profile of a vortex wall for a current pulse of $u$ = 50 m/s and $\beta$ = 0.05 with different current pulse widths. (f) Displacement profile of an antivortex wall at different current densities for $\beta$ = 0.05. Displacement profile of an antivortex for a current pulse of $u$ = 50 m/s and $\beta$ = 0.05 (g) and $\beta$ = 0 (h) at different current pulse widths.

We also studied vortex wall behavior under the current induced spin transfer torque. Even with zero nonadiabatic coefficients, we have observed a finite displacement of a vortex wall before its transformation to a transverse wall in figure 6(c) in the all current densities below the critical current ($u$ = 600 m/s for $\beta$ = 0). By increasing the nonadiabatic coefficient to 0.05, the forward displacement increases due to the presence of a nonadiabatic effective field in figure 6(d). The current pulse response of a vortex wall is similar to that of the field response such that by increasing of the length of the current pulse, the forward displacement of a vortex wall increases as seen in figure 6(e). Figure 6(f) shows the displacement profile of an antivortex wall due to the electric currents for the different current densities and $\beta$ = 0.05. Similar to the magnetic field excitation, a finite displacement of domain wall has been observed much below the critical current density. In addition, by increasing the external current density, the backward displacement of an antivortex wall decreases since the current enhances the transformation of the metastable wall into a stable wall.

The current pulse response of an antivortex wall is shown in figure 6(g). Depending on the pulse width, it is possible for an antivortex wall to displace in the either forward or backward direction till it transforms to a transverse wall. As mentioned before, the pinning sites act as a local harmonic potential that helps a conversion between the kinetic and potential energy. Therefore, it is expected that in a rough nanowire, even with $\beta$ = 0, one should observe the backward motion of an antivortex wall. As shown in figure 6(h), similar to the case with a finite nonadiabatic coefficient in figure 6(g), it is possible for an antivortex wall to have the either backward or forward motion depending on the current pulse width even though $\beta$ = 0.

In order to better understand the response of the vortex or antivortex wall to a current pulse, we have calculated the real time x- and y-positions of the domain wall as well as its tilting angle as shown in figure 7. For the vortex wall, we have injected a current pulse of 20 ns with $u$ = 50 m/s and $\beta$ = 0.05. As the vortex moves along the nanowire seen in the figure 7(a) and 7(b), its core displaces in the transverse direction. When the vortex core reaches the edge of the nanowire at 30.6 ns, the vortex wall transforms to a transverse wall. As shown in figure 7(c) there is a sudden drop in the tilting angle of the domain wall in 30.6 ns, which is a sign of the transformation from a vortex to a transverse wall. We also studied the real time behavior of an antivortex wall to a current pulse width of 0.1 ns with $u$ = 50 m/s and $\beta$ = 0. The x- and y-position of the antivortex wall show an oscillatory behavior as shown in figure 7(d) and 7(e) due to the presence of local harmonic potentials from the edge roughness. For the antivortex wall the transformation happens at 54.5 ns in which the antivortex core annihilates and a transverse wall nucleates in the nanowire. The tilting angle of an antivortex wall also presents a sudden drop from 90 degree to almost zero degree at 54.5 ns as shown in figure 7(f).

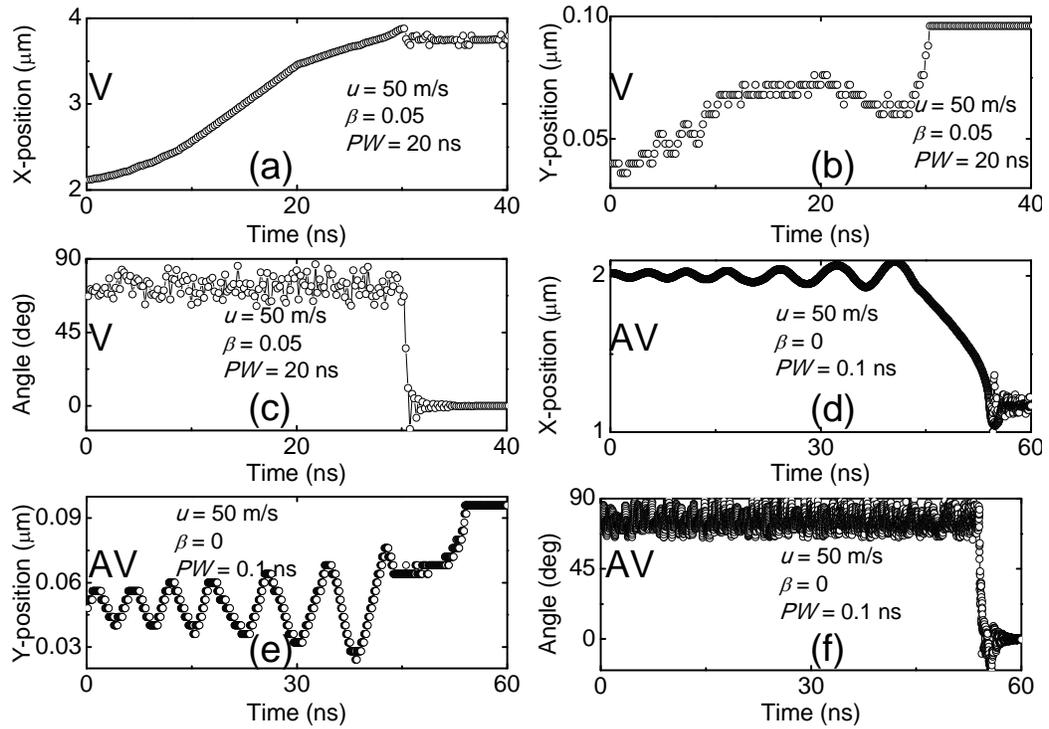

**Figure 7.** Time dependent position of a vortex wall in x- (a), y-direction (b), and tilting angle (c) for a current pulse of 20 ns with $u$ = 50 m/s and $\beta$ = 0.05. Time dependent position of an anitivortex wall in x- (d), y-direction (e), and tilting angle (f) for a current pulse of 0.1 ns with $u$ = 50 m/s and $\beta$ = 0.

*3.2. Random roughness*

We have also simulated the nanowire with a random roughness of $D$ = 4 nm as shown in the inset of figure 8. The displacement profile of an antivortex wall due to the current pulse excitation is demonstrated for a current of $u$ = 50 m/s and $\beta$ = 0.05. Since the antivortex depinning mechanism is similar to the nanowire with a periodic roughness in figure 6(g), the displacement profile strongly depends on the position of roughness and phase difference between the pulse width and antivortex oscillation, which enables bidirectional displacements.

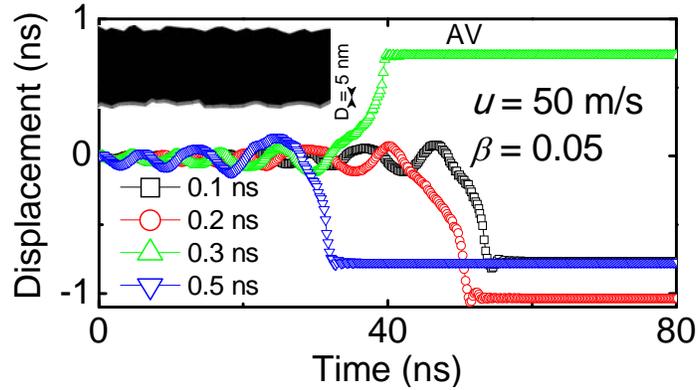

**Figure 8.** Displacement profile of an antivortex wall in a nanowire with random roughness for $u$ = 50 m/s and $\beta$ = 0.05.

## 4. Conclusion

We have studied the dynamics of metastable domain walls in a ferromagnetic nanowire. It is found that the average velocity of a metastable wall is smaller than a stable wall. In addition, depending on the structure of the metastable wall, it can displace in the either forward or backward direction for the same excitation. In a rough nanowire, in particular, the metastable wall could have a finite displacement in an excitation much below the critical field or current required to displace a stable wall. Our results demonstrate that the creation of a metastable wall could result in a displacement of domain wall below the critical field or current of a stable wall and also explain some of the previous experimental observations in which a bidirectional displacement as well as a very slow velocity of domain walls was reported.

**Acknowledgments**

This work is supported by the Singapore National Research Foundation under CRP Award No. NRF-CRP 4-2008-06.

**References**

[1] Hayashi M, Thomas L, Moriya R, Rettner C and Parkin S S P 2008 Current-controlled magnetic domain-wall nanowire shift register *Science* **320** 209-11

[2] Parkin S, Hayashi M and Thomas L 2008 Magnetic domain-wall racetrack memory *Science* **320** 190-4

[3] Barnes S E, Ieda J and Maekawa S 2006 Magnetic memory and current amplification devices using moving domain walls *Appl. Phys. Lett.* **89** 122507

[4] Allwood D A, Xiong G, Faulkner C C, Atkinson D, Petit D and Cowburn R P 2005 Magnetic Domain-Wall Logic *Science* **309** 1688-92

[5] Allwood D A, Xiong G, Cooke M D, Faulkner C C, Atkinson D, Vernier N and Cowburn R P 2002 Submicrometer Ferromagnetic NOT Gate and Shift Register *Science* **296** 2003-6

[6] McMichael R D and Donahue M J 1997 Head to head domain wall structures in thin magnetic strips *IEEE Trans. Magn.* **33** 4167-9

[7] Nakatani Y, Thiaville A and Miltat J 2005 Head-to-head domain walls in soft nano-strips: a refined phase diagram *J. Magn. Magn. Mater.* **290** 750-3

[8] Klaui M, Vaz C A F, Bland J A C, Heyderman L J, Nolting F, Pavlovska A, Bauer E, Cherifi S, Heun S and Locatelli A 2004 Head-to-head domain-wall phase diagram in mesoscopic ring magnets *Appl. Phys. Lett.* **85** 5637-9

[9] Shinjo T 2009 *Nanomagnetism and spintronics* (Amsterdam ; London: Elsevier)

[10] Laufenberg M, Backes D, Buhrer W, Bedau D, Klaui M, Rudiger U, Vaz C A F, Bland J A C, Heyderman L J, Nolting F, Cherifi S, Locatelli A, Belkhou R, Heun S and Bauer E 2006 Observation of thermally activated domain wall transformations *Appl. Phys. Lett.* **88** 052507

[11] Junginger F, Klaui M, Backes D, Rudiger U, Kasama T, Dunin-Borkowski R E, Heyderman L J, Vaz C A F and Bland J A C 2007 Spin torque and heating effects in current-induced domain wall motion probed by transmission electron microscopy *Appl. Phys. Lett.* **90** 132506-3

[12] Lacour D, Katine J A, Folks L, Block T, Childress J R, Carey M J and Gurney B A 2004 Experimental evidence of multiple stable locations for a domain wall trapped by a submicron notch *Appl. Phys. Lett.* **84** 1910-2

[13] Klaui M, Laufenberg M, Heyne L, Backes D, Rudiger U, Vaz C A F, Bland J A C, Heyderman L J, Cherifi S, Locatelli A, Mentes T O and Aballe L 2006 Current-induced vortex nucleation and annihilation in vortex domain walls *Appl. Phys. Lett.* **88** 232507

[14] Kläui M, Jubert P O, Allenspach R, Bischof A, Bland J A C, Faini G, Rüdiger U, Vaz C A F, Vila L and Vouille C 2005 Direct Observation of Domain-Wall Configurations Transformed by Spin Currents *Phys. Rev. Lett.* **95** 026601

[15] Hayashi M, Thomas L, Rettner C, Moriya R, Jiang X and Parkin S S P 2006 Dependence of current and field driven depinning of domain walls on their structure and chirality in permalloy nanowires *Phys. Rev. Lett.* **97** 207205

[16] Uhlig W C, Donahue M J, Pierce D T and Unguris J 2009 Direct imaging of current-driven domain walls in ferromagnetic nanostripes *J. App. Phys.* **105** 103902-7

[17] Nakatani Y, Shibata J, Tatara G, Kohno H, Thiaville A and Miltat J 2008 Nucleation and dynamics of magnetic vortices under spin-polarized current *Phys. Rev. B* **77** 014439

[18]	Heyne L, Rhensius J, Bisig A, Krzyk S, Punke P, Klaui M, Heyderman L J, Le Guyader L and Nolting F 2010 Direct observation of high velocity current induced domain wall motion *Appl. Phys. Lett.* **96** 032504-3
[19]	Li Z and Zhang S 2004 Domain-wall dynamics driven by adiabatic spin-transfer torques *Phys. Rev. B* **70** 024417
[20]	Zhang S and Li Z 2004 Roles of Nonequilibrium Conduction Electrons on the Magnetization Dynamics of Ferromagnets *Phys. Rev. Lett.* **93** 127204
[21]	Tatara G and Kohno H 2004 Theory of Current-Driven Domain Wall Motion: Spin Transfer versus Momentum Transfer *Phys. Rev. Lett.* **92** 086601
[22]	Tatara G, Kohno H and Shibata J 2008 Microscopic approach to current-driven domain wall dynamics *Phys. Rep.* **468** 213-301
[23]	Yamaguchi A, Ono T, Nasu S, Miyake K, Mibu K and Shinjo T 2004 Real-Space Observation of Current-Driven Domain Wall Motion in Submicron Magnetic Wires *Phys. Rev. Lett.* **92** 077205
[24]	Hayashi M, Thomas L, Bazaliy Y B, Rettner C, Moriya R, Jiang X and Parkin S S P 2006 Influence of Current on Field-Driven Domain Wall Motion in Permalloy Nanowires from Time Resolved Measurements of Anisotropic Magnetoresistance *Phys. Rev. Lett.* **96** 197207
[25]	Vanhaverbeke A, Bischof A and Allenspach R 2008 Control of Domain Wall Polarity by Current Pulses *Phys. Rev. Lett.* **101** 107202
[26]	Vernier N, Allwood D A, Atkinson D, Cooke M D and Cowbu R P 2004 Domain wall propagation in magnetic nanowires by spin-polarized current injection *Europhys. Lett.* **65** 526-32
[27]	Togawa Y, Kimura T, Harada K, Akashi T, Matsuda T, Tonomura A and Otani Y 2006 Current-excited magnetization dynamics in narrow ferromagnetic wires *Jpn. J. Appl. Phys.* **45** L683-L5
[28]	Chauleau J-Y, Weil R, euml, Thiaville A and Miltat J 2010 Magnetic domain walls displacement: Automotion versus spin-transfer torque *Phys. Rev. B* **82** 214414
[29]	Thomas L, Moriya R, Rettner C and Parkin S S P 2010 Dynamics of Magnetic Domain Walls Under Their Own Inertia *Science* **330** 1810-3
[30]	Schryer N L and Walker L R 1974 The motion of 180° domain walls in uniform dc magnetic fields *J. App. Phys.* **45** 5406-21
[31]	Nakatani Y, Thiaville A and Miltat J 2003 Faster magnetic walls in rough wires *Nat. Mater.* **2** 521-3
[32]	Seo S-M, Lee K-J, Yang H and Ono T 2009 Current-Induced Control of Spin-Wave Attenuation *Phys. Rev. Lett.* **102** 147202
[33]	Jamali M, Yang H and Lee K-J 2010 Spin wave assisted current induced magnetic domain wall motion *Appl. Phys. Lett.* **96** 242501-3
[34]	Choi S, Lee K-S, Guslienko K Y and Kim S-K 2007 Strong Radiation of Spin Waves by Core Reversal of a Magnetic Vortex and Their Wave Behaviors in Magnetic Nanowire Waveguides *Phys. Rev. Lett.* **98** 087205
[35]	Han D S, Kim S K, Lee J Y, Hermsdoerfer S J, Schultheiss H, Leven B and Hillebrands B 2009 Magnetic domain-wall motion by propagating spin waves *Appl. Phys. Lett.* **94** 112502
[36]	Seo S-M, Lee H-W, Kohno H and Lee K-J 2011 Magnetic vortex wall motion driven by spin waves *Appl. Phys. Lett.* **98** 012514-3
[37]	Malozemoff A P and Slonczewski J C 1979 *Magnetic domain walls in bubble materials* vol vii (New York: Academic Press)
[38]	Thiaville A, Nakatani Y, Miltat J and Suzuki Y 2005 Micromagnetic understanding of current-driven domain wall motion in patterned nanowires *Europhys. Lett.* **69** 990-6
[39]	Kronmüller H and Parkin S S P 2007 *Handbook of magnetism and advanced magnetic materials* (Chichester, West Sussex ; Hoboken, NJ: John Wiley & Sons)

[40]    Lee J-Y, Lee K-S, Choi S, Guslienko K Y and Kim S-K 2007 Dynamic transformations of the internal structure of a moving domain wall in magnetic nanostripes *Phys. Rev. B* **76** 184408

[41]    Gadbois J and Zhu J G 1995 Effect of edge roughness in nano-scale magnetic bar swtiching *IEEE Trans. Magn* **31** 3802-4

[42]    Martinez E, Lopez-Diaz L, Torres L, Tristan C and Alejos O 2007 Thermal effects in domain wall motion: Micromagnetic simulations and analytical model *Phys. Rev. B* **75** 174409

[43]    Tanigawa H, Koyama T, Bartkowiak M, Kasai S, Kobayashi K, Ono T and Nakatani Y 2008 Dynamical Pinning of a Domain Wall in a Magnetic Nanowire Induced by Walker Breakdown *Phys. Rev. Lett.* **101** 207203

[44]    Jiang X, Thomas L, Moriya R, Hayashi M, Bergman B, Rettner C and Parkin S S P 2010 Enhanced stochasticity of domain wall motion in magnetic racetracks due to dynamic pinning *Nat. Commun.* **1** 25

[45]    Thomas L, Hayashi M, Jiang X, Moriya R, Rettner C and Parkin S 2007 Resonant amplification of magnetic domain-wall motion by a train of current pulses *Science* **315** 1553-6

[46]    Martinez E, Lopez-Diaz L, Alejos O and Torres L 2008 Resonant domain wall depinning induced by oscillating spin-polarized currents in thin ferromagnetic strips *Phys. Rev. B* **77** 144417